# Towards Contactless Elevators with TinyML using CNN-based Person Detection and Keyword Spotting

Anway S. Pimpalkar[1] and Deeplaxmi V. Niture[1]

[1]COEP Technological University, Department of Electronics and Telecommunication, Pune, Maharashtra, India 411005

## Abstract

**Purpose:** This study introduces a proof of concept for a contactless elevator operation system designed to minimize human intervention while enhancing the safety, intelligence, and efficiency of user experiences in elevator systems.

**Design/methodology/approach:** A microcontroller-based edge device capable of executing tiny Machine Learning (tinyML) inferences is developed for elevator operation. The system employs person detection and keyword spotting algorithms based on tinyML, enabling the deployment of cost-effective, robust units with minimal infrastructural changes. The design integrates preprocessing steps and quantized convolutional neural networks within a multitenant framework to optimize accuracy and response time.

**Findings:** Empirical results demonstrate that the system achieves a person detection accuracy of 83.34% and keyword spotting efficacy of 80.5%, with an overall latency under 5 seconds. These metrics indicate the system's effectiveness in real-world scenarios.

**Originality:** Current contactless elevator technologies often face high deployment costs, limited viability, and inconsistent performance. Our proposed system, leveraging tinyML technologies, marks a significant paradigm shift in the elevator industry. This novel approach addresses existing challenges by providing a cost-effective, reliable, and scalable solution. The device has the potential to enhance user safety and operational efficiency without substantial infrastructural changes, contributing a valuable innovation to the field.

**Research limitations/implications:** While the study shows promising results, the scalability and integration with existing elevator systems require further exploration.

**Practical implications:** The demonstrated energy efficiency, operational simplicity, and enhanced safety suggest that the adoption of tinyML in elevator systems could revolutionize the industry, providing a model for future technological advancements in automated systems.

**Social implications:** By reducing the need for physical contact and improving the efficiency of elevator operations, this technology could significantly impact public health and user convenience in multi-floor buildings, especially in the context of pandemics or hygiene concerns.

# Introduction

The development of the elevator has played a crucial role in shaping the iconic cityscapes observed globally (Al-Kodmany, 2015, pp. 1070–1104; Gane and Haymaker, 2010, pp. 100–111). The construction of high-rises, which would have been impractical without elevators, has transformed urban landscapes. This industry's focus on enhancing speed, refining operating algorithms, and strengthening safety protocols represents the core of ongoing research and development efforts (Al-Kodmany, 2023, pp. 530–548). Amidst these priorities, the emergence of touch-free usage scenarios is notable, offering rapid interaction with minimal effort from users — a growing demand in contemporary society. Additionally, contactless systems offer hygienic advantages, particularly in devices situated in public spaces. In the context of elevators, the adoption of contactless operation, driven by these factors, presents an innovative avenue for implementation (Bo and Zheyi, 2020, pp. 311–315).

## Current Contactless Elevator Solutions

To integrate contactless functionalities into elevator systems, a wide array of innovative technologies is being explored. Work has been done demonstrating how contactless operation of elevators can be achieved using infrared sensors (Rossi *et al.*, 2021, pp. 179–185; Vaish *et al.*, 2020, pp. 1–6), passive RFID tags (Misran *et al.*, 2014, pp. 252–257; Xuefu Xiao *et al.*, 2011, pp. 126–129), hand-gesture controls (Lai *et al.*, 2022, p. 411), and QR codes (Reinsalu and Robal, 2023, pp. 1–4; Rubies *et al.*, 2023, p. 3971), Bluetooth connectivity (Cheng Jing and Guo-jun Zhao, 2014, pp. 350–355; Hangli *et al.*, 2018, pp. 566–570), and smartphone applications (Basmaji *et al.*, 2021, pp. 1–5; Liu *et al.*, 2022, p. 2523). These technological advancements are aimed at minimizing physical interaction with elevator controls, a significant step towards enhancing hygiene and user convenience. However, these technologies are accompanied by numerous challenges. Infrared sensors may experience false activations or fail to detect presence under fluctuating lighting conditions or when obstructions occur (Özcan *et al.*, 2020, p. 107633). Additionally, they lack the capability to differentiate between deliberate user interactions and inadvertent movements. The reliability of hand-gesture controls may be compromised by variations in individual users' gestures, often necessitating a period of adaptation. Moreover, Bluetooth connectivity demands the possession of a Bluetooth-capable device by users and may introduce privacy and security risks (Ibn Minar, 2012, pp. 127–148). Connectivity problems are particularly prevalent in environments with significant electronic interference (Doufexi *et al.*, 2003, pp. 680–684 vol.1).

Voice recognition technologies have become more promising compared to touch interfaces because they allow for hands-free interaction, significantly reducing the need for physical contact with devices and thus the potential spread of viruses. Although speaking can indeed emit respiratory droplets, voice recognition technology typically works effectively at a distance, allowing users to interact from a safe, virus-transmission-reducing range (Basmaji *et al.*, 2021, pp. 1–5). They provide an effective solution not just for improving hygiene but also for ensuring accessibility for people with disabilities. By enabling users to operate elevators through voice commands, these systems remove physical barriers and enhance usability for all (Shinde *et al.*, 2021, pp. 728–733). These contactless technologies are being gradually adopted in various regions, with many already in use in advanced elevator systems. Their implementation represents a significant leap forward in making elevator use safer, more efficient, and more inclusive in both public and private buildings.

## *Need for New Solutions*

Present elevator systems typically come equipped with dedicated computing resources. Retrofitting these to accommodate new contactless technologies could prove costly for standard housing and office spaces in typical communities. Scaling this system for every elevator on each floor can rapidly increase the overall costs. Consequently, there is a need for affordable solutions that necessitate minimal alterations to the existing infrastructure. Ideally, these solutions should seamlessly integrate with the elevator's control buses, offering a cost-effective way to upgrade to contactless operation without the need for extensive hardware overhauls.

Beyond the aspect of cost, operating multiple algorithms on a central mainframe computer necessitates a system with exceptionally high computational power, capable of running numerous inferences simultaneously. Such a setup poses challenges in terms of scalability. An alternative strategy involves decentralizing computations to individual devices via edge computing. By executing inferences on edge devices, specifically the microcontrollers located at each floor unit, significant latency reductions are achieved. This approach negates the need for data transmission to a central processing unit, enabling faster and more efficient response to user commands. This decentralized, edge-based architecture serves not only to conserve computational resources but also to enhance the overall responsiveness and reliability of the system (Zhang *et al.*, 2018, pp. 39–45).

A field of interest which can serve as a starting point in this space is tinyML (David *et al.*, 2020). It offers a viable solution for efficiently running machine learning models on low-power, resource-constrained devices. This technology is particularly suitable for implementing advanced features, like those required for contactless elevator systems, without the need for high-powered computing resources. The study introduces a novel contactless elevator system that harnesses tinyML technologies, incorporating edge microcontrollers to create an economical solution compared to conventional systems. These microcontrollers deliver essential functionalities at substantially lower costs, enabling the widespread commercial implementation of contactless technologies across diverse socio-economic strata. However, this cost-efficiency comes with its own set of design challenges. The edge devices used typically possess limited memory and computational capabilities, presenting hurdles in the deployment of full-scale algorithms (Kallimani *et al.*, 2023).

In any industrial product development, user expectations regarding robustness, accuracy, and latency are critical factors that dictate viability (Sireli *et al.*, 2007, pp. 380–390). The concept proposed in this study examines the deployment of tinyML to significantly advance the contactless operation of elevator systems keeping these expectations in mind. This concept envisions a system that incorporates person detection and voice command features to facilitate a contactless interface, aiming to meet the high standards expected by users in today's technologically advanced landscape. Positioned outside the elevator on each floor as shown in Figure 1, replacing conventional button interfaces, the system operates through three distinct phases:

1. **Person Detection:** This phase involves identifying a passenger waiting in front of the elevator. The detection unit is designed to recognize the presence of a person intending to use the elevator.

2. **Keyword Spotting:** Once a person is detected, the system enters the keyword spotting phase. Here, it listens for the passenger to vocalize a specific floor number, indicating their desired destination.
3. **Response System:** After successfully detecting the spoken floor number, the system responds by communicating this destination to the central elevator control. This process directs the elevator to the correct floor as per the passenger's request.

These steps are visually represented in Figure 2, which outlines the overall block diagram of the working methodology and the actions involved in the proposed elevator system. This diagram illustrates the flow of information and control from person detection to the final elevator response. The design of the contactless elevator system incorporates two primary pipelines, with each unit on every floor executing these pipelines to facilitate the steps of person detection, keyword spotting, and response generation. The methodology central to this system utilizes convolutional neural networks (CNNs) for performing inferences in both pipelines (Janapa Reddi *et al.*, 2022).

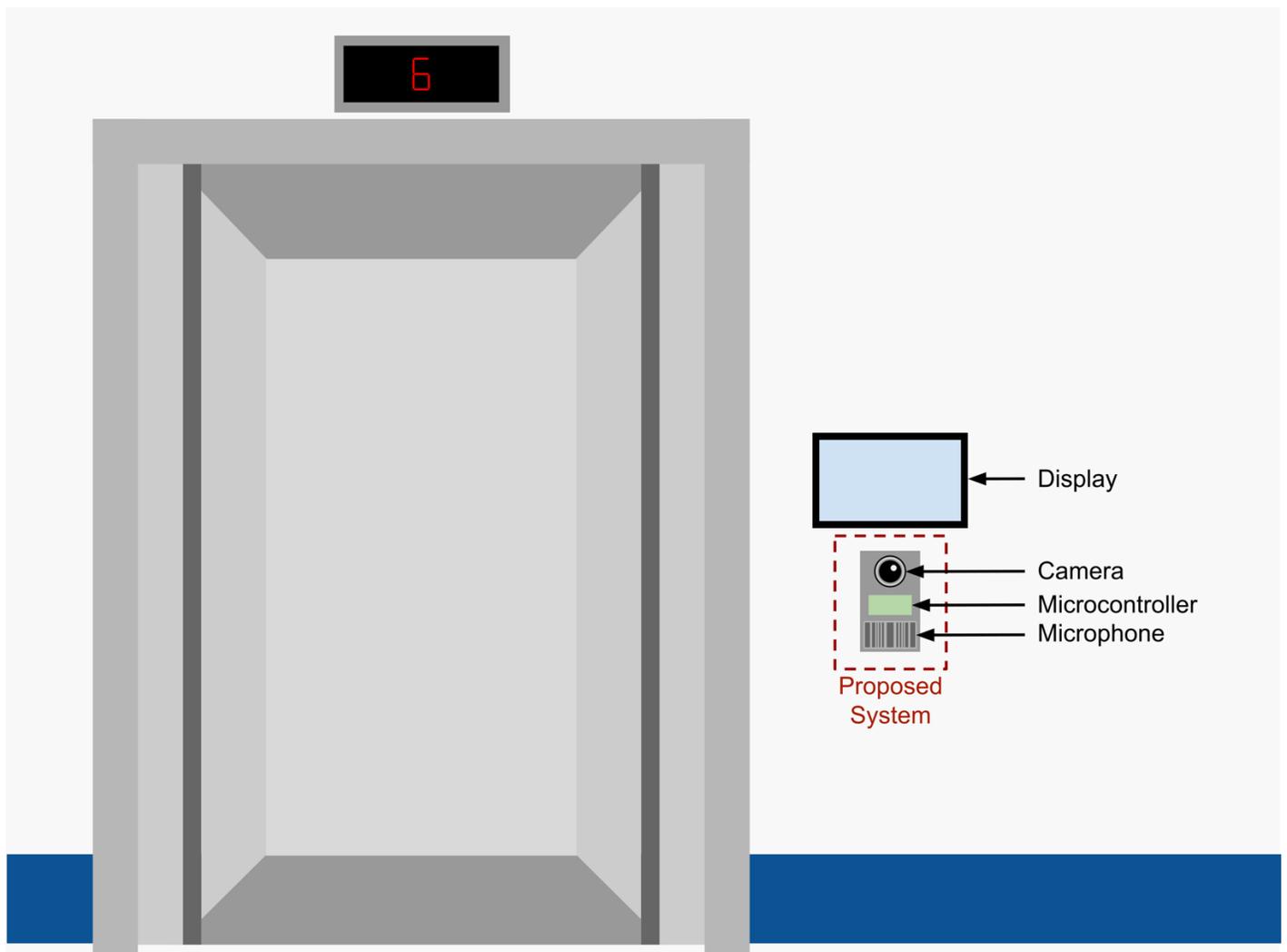

*Figure 1. Deployment of the built device outside an elevator.*

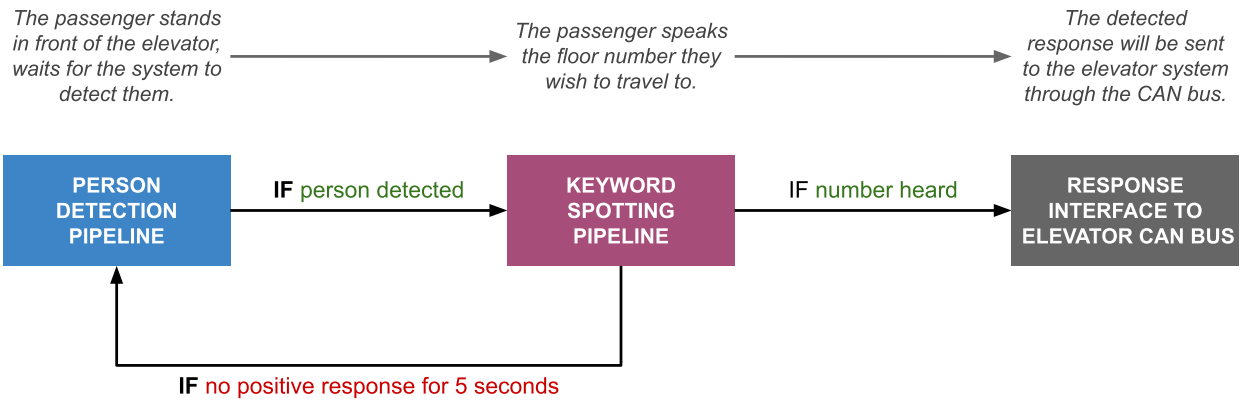

*Figure 2. Overall block diagram of working methodology and actions involved in the proposed elevator system.*

# System Construction and Algorithm Pipelines

## Device Construction

For this work, a selection of readily available components, including a microcontroller, camera, and microphone, interfaced as shown in Figure 3. These design choices are made with the intention of minimizing costs within the project's scope, thereby serving as a proof of concept. This approach demonstrates how even basic, widely accessible devices can effectively perform the tasks required for a sophisticated contactless elevator system. The Arduino Nano 33 BLE Sense, (Arduino LLC, MA, USA) serves as the central microcontroller development board. It boasts 256 KB of SRAM and a comprehensive array of sensors, including an onboard microphone utilized for the keyword spotting functionality. Image data necessary for person detection is captured using the OV7675 Camera Module (Arducam, Nanjing, China). For a practical demonstration, this system is applied to a hypothetical elevator in a building with four accessible floors, establishing a focused proof of concept.

The system is structured into two main subsystems, each characterized by its own sequence of operational steps, as illustrated in Figure 4. The outcome of the second pipeline's process is then conveyed to the central elevator system.

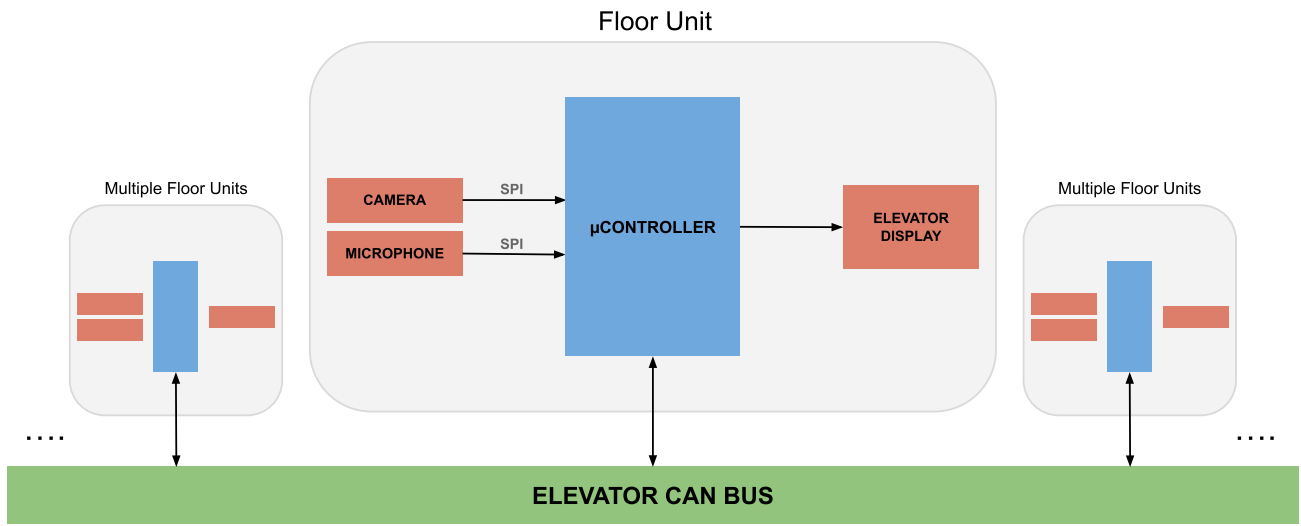

*Figure 3. Hardware components and interfaces.*

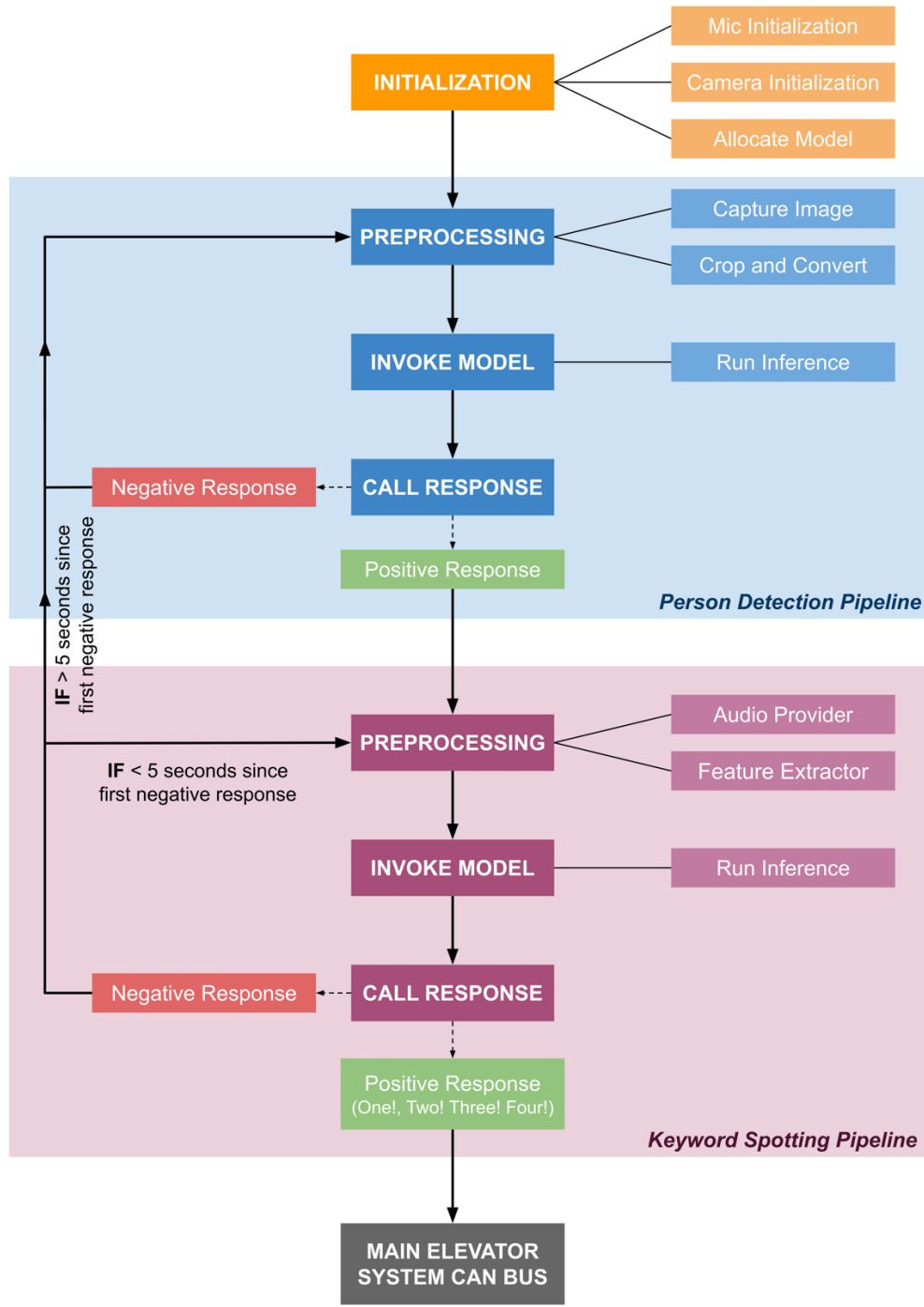

Figure 4. Detailed steps involved in the Person Detection and Keyword Spotting pipelines.

## Person Detection Pipeline

Detecting a person in an image frame captured by the elevator unit is the first step in enabling the contactless elevator. The pipeline can be divided into four stages – initialization, preprocessing, invoking the model, and the call response.

*Initialization* – In the initial phase of the contactless elevator's operation, the OV7675 camera module is configured to capture images at a rate of 5 fps in grayscale mode, a choice driven by the lower memory demands of grayscale imaging. Concurrently, the convolutional neural network (CNN) model is prepared for execution on the device, detailed in the next section. This preparation involves allocating memory for the model, defining its input parameters, initializing the interpreter, and establishing the model's main operational loop, all of which are critical for the seamless functioning of the system (David *et al.*, 2020).

*Preprocessing* – Next, a data frame is captured using the OV7675 module and converted to a format readable by our neural network, i.e., 96 × 96 px grayscale.

*Invoking the Model* – The preprocessed image is run through a trained CNN to attain an inference. The network outputs two binary classes indicating whether a person is detected in the frame, which further triggers one of two call responses.

*Call Response* – If the camera detects no person waiting for the elevator, then a negative call response is initiated, and the pipeline loops back to the preprocessing step of the pipeline. Alternatively, if a person is detected, the keyword spotting pipeline preprocessing step is invoked to listen to the person speak.

## Keyword Spotting Pipeline

The system begins listening for keywords if a person is detected standing before the elevator. Similar to the previous pipeline, it can be divided into four main stages – initialization, preprocessing, invoking the model, and the call response.

*Initialization* – First, the microphone built onto the Arduino Nano 33 BLE Sense is initialized. The CNN model space is allocated to the keyword spotting model. This step is completed only once.

*Preprocessing* – In the audio provider stage of the system, the microphone is set up to capture audio data over a specific timeframe. This stage involves converting the incoming analog audio signals into a digital format for further processing. The audio data is sampled at a rate of 16 kHz. Once converted, the digital audio data is continuously loaded into an audio buffer. This buffer plays a pivotal role in the system's ability to detect when a user utters a specific keyword. The audio samples are stored as 16-bit values within this buffer, providing a sufficient resolution for the subsequent feature extraction stage.

To facilitate efficient processing and classification of audio signals by the CNN, feature extraction from the audio buffer sample is crucial before it enters the inference pipeline. Initially, a 30ms segment of the audio buffer is captured, and a Fast Fourier Transform (FFT) is applied to it, resulting in 256 values. These values are then averaged to create a more robust representation, reducing them to 43 values. These 43 values correspond to a single slice of the spectrogram that the system aims to construct. Each FFT essentially forms a single column in the complete spectrogram of the audio buffer. A full spectrogram is generated once data equivalent to one second has been compiled. This process effectively transforms the time series audio signal into a frequency domain signal, where further optimizations are possible. One such optimization is the generation of Mel Frequency Cepstral Coefficients (MFCCs), which are particularly adept at capturing features not easily discernible to the human ear. Since human hearing is more sensitive to lower frequencies, focusing on these frequencies can enhance keyword detection (Muda *et al.*, 2010). The MFCCs place greater emphasis on these lower frequencies, enabling the machine learning model to more accurately identify the desired keywords (Abdul and Al-Talabani, 2022, pp. 122136–122158). These MFCCs are then fed into the CNN model as inputs for inference, optimizing the system's ability to recognize and process voice commands efficiently and accurately. Example visualizations of the MFCCs used in this implementation are shown in Figure 5. Each of the four MFCC visualizations presents a distinct pattern, reflective of the unique audio characteristics: "One" suggests a sound with consistent, low-frequency energy; "Two" indicates a complex sound with periodicity; "Three" shows mid-range frequency variations; and "Four" represents a dynamic sound with a spread of energy across mid to high frequencies. These differences in energy distribution and consistency across the frequency bands imply variations in pitch, complexity, and temporal changes inherent to each audio sample.

*Invoking the Model* – The MFCC spectrogram is run through trained CNN, which assigns one of six classes to each captured spectrogram – "One," "Two," "Three," "Four," "Unknown," or "Silence."

*Call Response* – If the inference is returned to be positively identified as a given floor number (one, two, three, or four), the pipeline feeds the floor number as input to the central elevator system. If no input is found (unknown or silence), the keyword spotting pipeline is looped back to its preprocessing step. If nothing is heard for 5 seconds, the keyword spotting loop is exited, and the person detection pipeline is executed from the preprocessing stage.

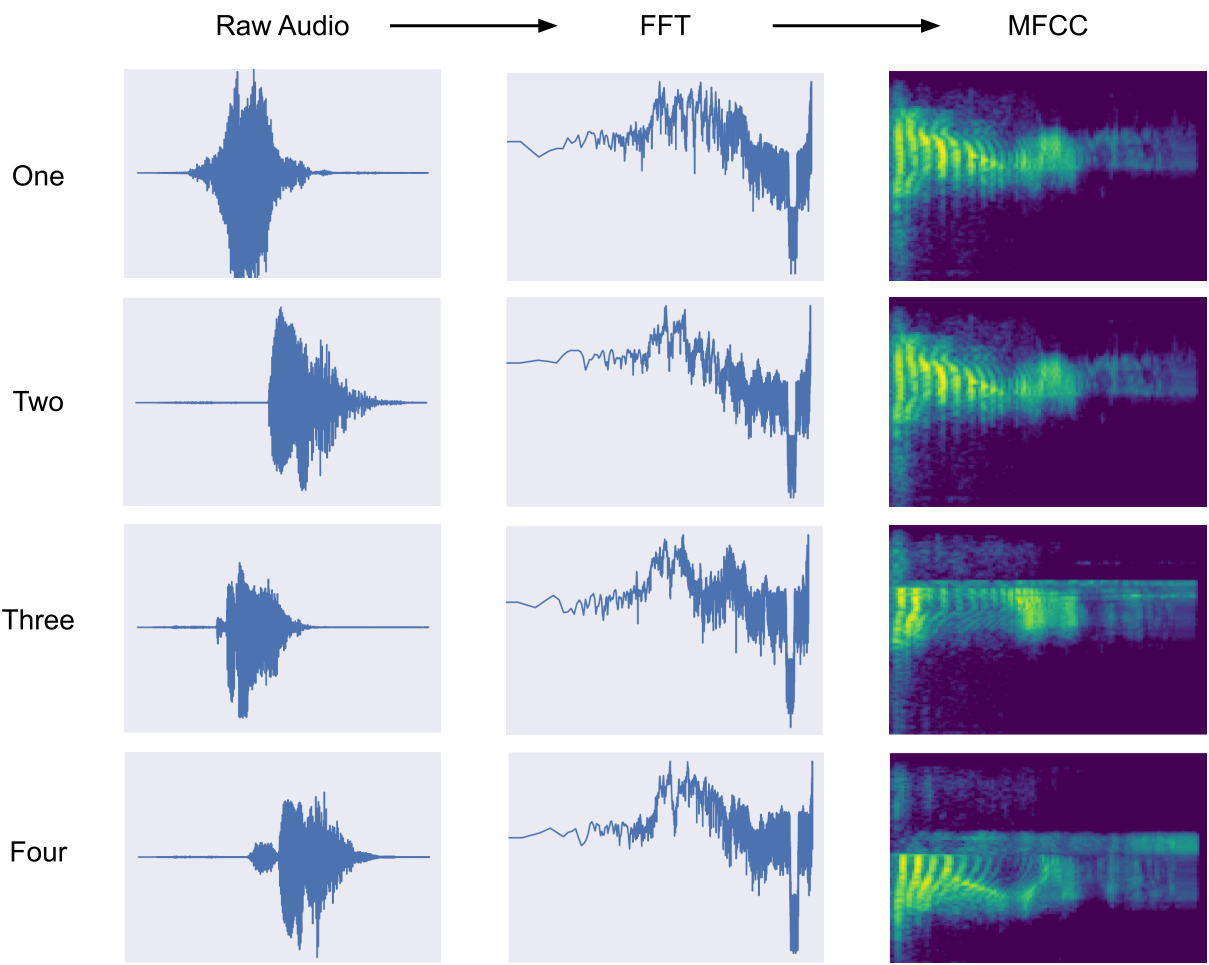

*Figure 5. Sample raw audio, FFT, and MFCC representations for spoken "One", "Two", "Three", and "Four".*

## Response System

All existing elevator technologies today use a CAN bus to coordinate between different system elements (Salunkhe *et al.*, 2016, pp. 301–304; Wan *et al.*, 2002, pp. 98+100-98+103). Integrating the output from the keyword spotting pipeline into the main bus of both existing and new elevator systems involves customizing the unit based on the specifications provided by the manufacturer. For the current implementation, this process requires developing an interface module that can translate the voice command signals into a format compatible with the elevator's control system. For existing systems, this might mean creating an adapter that can interpret the voice-activated commands and convert them into the electrical signals traditionally generated by button presses. For new systems, manufacturers might design the control bus with built-in compatibility for such voice command interfaces. Additionally, secure and reliable communication protocols need to be established to ensure that the voice commands are accurately and consistently interpreted by the elevator system. This could involve the use of advanced encryption and error-checking algorithms to prevent misinterpretation and unauthorized access. By carefully designing this integration, the system can seamlessly incorporate voice commands as a natural extension of the elevator's control mechanisms, enhancing user experience and accessibility (Jian Chu and Qing Lu, 2012, pp. 820–823). The detailed integration of this system with an actual elevator control data bus, including specific protocols and interfaces, falls outside the scope of this study.

## CNN and TinyML Implementations

The previous section highlighted how convolutional neural networks (CNNs) are employed for inferencing in both pipelines of the system, operating on the microcontroller. Adhering to tinyML principles, each model in the proposed system has been independently trained to meet its specific objective within the resource-limited environment of a microcontroller.

*Person Detection Model*

The person detection model for the system utilizes the *MobileNetV1* architecture (Howard *et al.*, 2017), selected for its suitability in a constrained environment like a microcontroller. Post-training, the model undergoes quantization to further optimize it for the limited resources available (Heim *et al.*, 2021). *MobileNetV1* is chosen over its successor, *MobileNetV2*, primarily because it demands less RAM for intermediate activation buffers during runtime, making it more appropriate for applications with stringent memory limitations (Sandler *et al.*, 2018, pp. 4510–4520). Additionally, the implementation employs a reduced depth multiplier in the *MobileNetV1* architecture (Howard *et al.*, 2017), ensuring that the model's size remains within the 250KB flash memory storage capacity of the microcontroller. This careful balance of model complexity and memory requirements is crucial for maintaining effective performance within the constraints of the hardware.

Deploying computer vision models to microcontroller environments requires minimal energy and memory footprints. To facilitate the development of such models, Chowdhery et al. constructed a new dataset called Visual Wake Words for binary classification indicating the presence of a person (Chowdhery *et al.*, 2019). A total of 115,000 images are sourced from the Microsoft Common Objects in Context (MS COCO) dataset (Lin *et al.*, 2014, pp. 740–755) and relabeled as 'Person' or 'No-person' as a part of the new dataset.

During the training of the person detection model, data augmentation techniques such as horizontal shifts of 10%, vertical shifts of 10%, rotation in the range of 90°, horizontal flips and vertical flips have been applied to enhance the model's ability to accurately detect a person under various conditions. The images used for training are resized to 96 × 96 pixels and converted to grayscale to match the capabilities of the OV7675 camera module and are used for the training process. To optimize the neural network's performance on edge devices, the model weights are frozen and quantized, which optimizes for low memory performance, latency, and compatibility with hardware devices that do not support floating point calculations (David *et al.*, 2020).

*Keyword Spotting Model*

To identify spoken keywords, the system utilizes an architecture resembling that built by Sainath et al. for keyword spotting applications (Sainath and Parada, 2015). The *tiny_conv* model architecture, trained to leverage its pre-optimized design for deployment on embedded microcontrollers. This architecture is lightweight, consisting of a depthwise separable convolutional layer and a fully connected layer, culminating in a softmax layer for output normalization (Bayerl *et al.*, 2020, pp. 460–465).

The crowdsourced Speech Commands dataset (Warden, 2018) is used to train the *tiny_conv* architecture. It contains many everyday speech commands required by keyword spotting applications, including the ones needed for the given classifier (Bayerl *et al.*, 2020, pp. 460–465). The number of keyword instances for "one," "two," "three," and "four" in the Speech Commands dataset are 3890, 3880, 3727, and 3728 respectively.

*Multitenant Architecture*

Memory and energy constraints on microcontrollers such as the Arduino Nano 33 BLE Sense makes running two deep learning simultaneous inference engines a challenge. Implementing a multitenant architecture for both models optimizes the deployment. Here, both models are allocated the same memory, but an inference is run by only one of the two models at a given time. This enables higher inference speeds and efficient memory allocation at runtime. In such an architecture, synchronization of data streams from the camera and microphone is also not required, and hence both can run at different sampling rates (David *et al.*, 2020).

## Results and Discussion

The Arduino Nano 33 BLE Sense and OV7675 module are mounted on a shield for ease of use and testing.

*Inferences*

After correctly positioning the shield, the system's effectiveness has been tested by conducting inferences with a subject standing both inside and outside the camera's frame. The inference process generates a probability value ranging from -127 to +127 (7-bit predictions), which are then further converted to a percentage value. Values closer to +127 indicate a higher likelihood of a human presence in the captured frame. The threshold is set at 59% through a trial-and-error process, optimizing for the latency and robustness. It also presents sample grayscale images that include a human subject as captured by the camera. When these images are processed through the model's algorithms, the system yields a positive output, indicating human presence. Despite the images being somewhat distorted and incomplete, the algorithm successfully identifies human features. This capability stems from the extensive training parameters and dataset augmentation used. The robust training ensures that even under less-than-ideal conditions, the system can accurately discern human presence, highlighting its efficiency and reliability in real-world applications.

RGB status indicators on the Arduino are used to notify the user, as shown in Figure 6. In future implementations, a sophisticated LCD/LED display can be used. The keyword spotting model inference pipeline is called after a positive callback from the person detection pipeline.

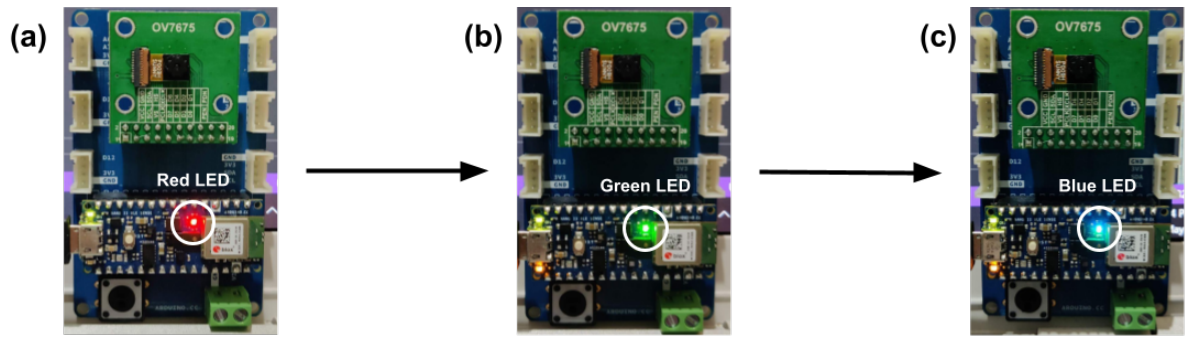

*Figure 6. LED indicators for the person detection and keyword spotting tasks: (a) Default red LED indicates that no person is detected by the device, (b) Green LED indicates that a person has been spotted and the device starts listening to the user's voice commands, and (c) Blue LED indicates that a floor number has been detected by the device.*

### Accuracy and Latency

Both models trained yielded good accuracies for both, pre and post quantization, as shown in Table I. During the hardware implementation testing, the model proved to be robust and quick to respond to both visual wake words and audio keywords. The model shows a latency low enough to be feasible as a practical product. Using the average of one hundred instances, the inference of the person detection model takes approximately 740 milliseconds, and an inference of the speech recognition model takes about 30 milliseconds. It is possible to call an elevator and record the destination floor number in approximately 4 to 5 seconds with the current prototype.

*Table I: Summary of Training Results for CNN Models*

| Model | Accuracy | Loss |
|---|---|---|
| Person Detection (MobileNetV1) | 83.89% | 0.42 |
| Person Detection (MobileNetV1) – *Quantized* | 80.34% | 0.53 |
| Keyword Spotting (tiny_conv) | 87.83% | 0.32 |
| Keyword Spotting (tiny_conv) – *Quantized* | 83.50% | 0.43 |

## Conclusion

The current proof of concept demonstrates acceptable accuracy and latency levels, yet these aspects may require further refinement for deployment in industrial settings. Balancing these two parameters can be achieved by modifying the number of nodes in the inference network and experimenting with different microcontrollers, each offering varying specifications and memory capacities. For instance, the SparkFun Edge board (SparkFun Electronics, Niwot, Colorado, USA) with its 384 KB SRAM, offers 50% more memory than the Arduino Nano 33 BLE Sense used in this work. However, it operates at a lower clock speed of 48 MHz compared to the Arduino's 64 MHz. These trade-offs could result in higher accuracy, albeit at the cost of marginally increased latency, but they can be optimized for specific use cases. Advanced edge devices equipped with integrated GPUs are also viable options, although they come with higher costs.

Several challenges need addressing in future iterations of this system. For instance, the presence of multiple people simultaneously speaking floor numbers near the elevator could lead to confusion in the inference pipelines. Additionally, ambient noise prevalent in public spaces like malls or schools must be factored into the model's training to ensure robust performance.

One significant future direction for this work involves the integration with the CAN bus of existing elevator systems. By this, the system can transmit the processed commands from the microcontroller directly to the elevator's operational controls. This approach would streamline the installation process in existing structures, making the upgrade to contactless technology more seamless and less intrusive. Additionally, exploring the compatibility of the system with different types of elevator models and brands is crucial. Each elevator system may have unique communication protocols and operational nuances. Developing a universal or adaptable interface that can be customized to different elevator systems could significantly expand the market reach and applicability of the technology.

Future enhancements of the system could also incorporate additional functionalities. For example, counting the number of people waiting for the elevator could help optimize its operational algorithm. Integrating fire safety mechanisms, force transducers, and speed sensors within the floor units and the elevator car could provide valuable information to waiting passengers, enhancing both safety and user experience. Such developments would not only improve the system's functionality but also its practicality and appeal in real-world applications.

## Video Demonstration and Code Availability

A supplementary video demonstrating the system's usage has been attached to this manuscript. Additionally, the project's source code is available on GitHub at https://github.com/anwaypimpalkar/smartElevatorSystem.